\documentclass{elsart3p}
\usepackage{amssymb, url}
\usepackage{graphicx}
\newcommand{\uu}[1]{\, \mathrm{#1}}

\begin{document}

\begin{frontmatter}
\title{ Commissioning Run of the CRESST-II Dark
Matter Search}

\author[mp]{G.~Angloher}, \author[tb]{M.~Bauer},
\author[mp]{I.~Bavykina},  \author[mp,co]{A.~Bento},
\author[ox]{A.~Brown}, \author[gs]{C.~Bucci},
\author[tu]{C.~Ciemniak}, \author[tu]{C.~Coppi},
\author[tb]{G.~Deuter}, \author[tu]{F.~von~Feilitzsch},
\author[mp]{D.~Hauff}, \author[ox]{S.~Henry}, \author[mp]{P.~Huff},
\author[ox]{J.~Imber}, \author[ox]{S.~Ingleby},
\author[tu]{C.~Isaila}, \author[tb]{J.~Jochum},
\author[mp]{M.~Kiefer}, \author[tb]{M.~Kimmerle},
\author[ox]{H.~Kraus}, \author[tu]{J.-C.~Lanfranchi},
\author[mp]{R.~F.~Lang}, \author[ox,bl]{B.~Majorovits},
\author[ox]{M.~Malek}, \author[ox]{R.~McGowan},
\author[ox]{V.~B.~Mikhailik}, \author[mp]{E.~Pantic},
\author[mp]{F.~Petricca}, \author[tu]{S.~Pfister},
\author[tu]{W.~Potzel}, \author[mp]{F.~Pr\"obst},
\author[tu,qu]{W.~Rau}, \author[tu]{S.~Roth},
\author[tb]{K.~Rottler},
\author[tb]{C.~Sailer}, \author[mp]{K.~Sch\"affner},
\author[mp]{J.~Schmaler}, \author[tb]{S.~Scholl},
\author[mp]{W.~Seidel},  \author[mp]{L.~Stodolsky},
\author[ox]{A.~J.~B.~Tolhurst}, \author[tb]{I.~Usherov} and
\author[tu,de]{W.~Westphal}

\address[mp]{Max-Planck-Institut f\"ur Physik, F\"ohringer Ring 6,
D-80805 M\"unchen, Germany}
\address[tu]{Physik-Department E15, Technische Universit\"at
M\"unchen, D-85747 Garching, Germany}
\address[ox]{Department of Physics, University of Oxford, Oxford
OX1 3RH, United Kingdom}
\address[tb]{Eberhard-Karls-Universit\"at T\"ubingen, D-72076
T\"ubingen, Germany}
\address[gs]{INFN, Laboratori Nazionali del Gran Sasso, I-67010
Assergi, Italy}
\address[bl]{now at: Max-Planck-Institut f\"ur Physik, F\"ohringer
Ring 6, D-80805 M\"unchen, Germany }
\address[co]{on leave from: Departamento de Fisica, Universidade de
Coimbra, P3004 516 Coimbra, Portugal}
\address[qu]{now at: Department of Physics, Queen's University,
Ontario K7L 3N6, Canada}
\address[de]{Deceased}

\begin{abstract}

The CRESST cryogenic direct dark matter search at  Gran Sasso,
searching for WIMPs
via nuclear recoil,
has been upgraded to CRESST-II by  several changes and
improvements. The upgrade includes 
a new detector support structure capable of   accommodating 33 
modules,  the
associated multichannel readout with 66 SQUID channels,  a neutron
shield, a calibration source lift, and the installation of a muon
veto. We present the
results of a commissioning run carried out in 2007.

 The basic element  of
CRESST-II  is a  detector module consisting of a large ($\sim$300
g)
 $\mathrm{CaWO_4}$ crystal and a very sensitive smaller ($\sim 2$
g)
light  detector to detect the
  scintillation light  from the $\mathrm{CaWO_4}$.
The large crystal gives an  accurate total energy 
measurement.  The  light  detector permits a determination 
 of the light  yield for an event,  allowing an effective
separation of nuclear recoils
from electron-photon backgrounds. Furthermore, information
from light-quenching factor studies allows the definition of a
region of the energy-light yield plane which  corresponds to
tungsten recoils. A neutron test is reported which supports the
principle of
using the light yield to identify the recoiling nucleus.

Data obtained   
 with two detector modules for a total exposure of 48 kg-days are
presented. Judging by the rate of events in the ``all nuclear
recoils'' acceptance region
the apparatus  shows a  factor $\sim$ten improvement
 with respect to  previous
results,
which we attribute  principally to the presence of the
neutron shield.
In the ``tungsten recoils''  acceptance region  three 
events are found, corresponding to a rate of 0.063 per kg-day. 
  Standard assumptions on the dark matter flux,  coherent or 
spin independent interactions,  
  then yield  a limit  for
WIMP-nucleon scattering of $4.8 \times 10^{-7}\uu{pb}$, at
$M_{\uu{WIMP}}\sim50\uu{GeV}$.

\end{abstract}
\end{frontmatter}

\section{Introduction}
Evidence continues to grow that the majority of the mass of our
galaxy and in the universe is made up of non-baryonic dark matter
 \cite{bertone2005}. Precision measurements of the cosmic microwave
background have given
 accurate figures for the density of matter and energy in the
universe as
 a whole, suggesting that about a fourth of the
 energy density of the universe is in the form of  dark matter
\cite{komatsu2008}.
 Measurements of the rotation curves of other spiral galaxies,
indicate large
 amounts of dark matter, which would  suggest it also  dominates 
our
 own galaxy. Gravitational  lensing of light by galactic clusters 
indicates large amounts of dark matter, and dark matter is
necessary
for a reasonable
description of the formation of galaxies.
 
 Although some of these observations might also be explained by
alternative theories
such as modified Newtonian dynamics
 (MOND) \cite{sanders2002}, recent studies of the Bullet Cluster
appear to
  favour the dark matter hypothesis 
 \cite{clowe2006}. However  the need  for direct detection
of  dark matter evidently remains strong.

A plausible origin for dark matter comes from particle physics in
the form of WIMPs (weakly interacting massive particles). These
would be
  stable massive particles with an interaction cross section
typical for weak interaction processes. A relic density 
 sufficient to make a significant contribution to the energy
density of the universe
 arises naturally  in this way.
 Supersymmetry provides  a well motivated
candidate in the form of the  neutralino and this has been
extensively
studied theoretically.

According to  the WIMP hypothesis  a galaxy has  a large 
gravitationally
bound  halo of such 
particles, making up most of the mass.  The   local mass  
density
may be   estimated  from halo models and is found to 
be  on the order of  $0.3\uu{GeVcm^{-3}}$. The range of the
coherent or spin-independent
 WIMP-nucleon scattering cross-sections predicted by minimal
supersymmetric models
 extends from below $10^{-10}$ to above $10^{-7}\uu{pb}$
\cite{trotta2007,trotta2007b};
 this makes the direct detection of such particles a difficult but
in  principle achievable task.

 The CRESST  project is a dark matter search aiming to detect the
nuclear scattering of
 WIMPs by the use of  cryogenic detectors. Since only small  
 recoil energies ($\sim$ keV's)  are   anticipated in  WIMP-nucleus
scattering, cryogenic detectors with their high sensitivity are
well
suited to the problem.
 
   CRESST-II is an upgrade of the original CRESST appartus
where the detectors are arranged in modules, each  consisting of
two
detectors, a large detector
 comprising the target mass  and a similar
but much smaller light
 detector measuring the light yield. While the large detector
gives  an accurate  total  energy measurement,
 the light yield  measurement   permits an effective rejection 
of  events which are not  nuclear recoils.  This arises
from the property of  electron-photon events,  constituting the
dominant
background, to give a much higher  light output than nuclear
recoils.

 The large detector consists of a $\mathrm{CaWO_4}$
crystal of  $\sim300\uu{g}$ with a
 tungsten superconducting-to-normal thermometer deposited on the
surface.
 The energy of an event  is measured in this
detector. The energy deposited
in the crystal is quickly converted into a gas of phonons  which
are
then absorbed in the superconducting thermometer.
 The  separate  light detector, based on the same principle, 
measures the  
   simultaneously emitted   scintillation light.     
 A comparison of these two signals then allows  an effective
background discrimination.

 Results from an earlier   prototype run with two modules 
were presented in Ref.\,\cite{angloher2005}, where taking  the
better of the two
modules and assuming  coherent
WIMP scattering on the tungsten
 led to the  limit $1.6\times10^{-6}\uu{pb}$.
 The improvements
discussed here are aimed at 
 ultimately  increasing this sensitivity
 by  up to two orders of magnitude. Such an increase would result,
for example, with background free data and a threshold of 12 keV
from 10 kg for 100 days giving 1000 kg-days. Or if the  threshold
could be reduced to 5 keV, with 300 kg days.

During 2007 an extended commissioning run of the new  setup
  was carried out.
 Although this was
 primarily for optimization purposes, the results also represent
  an interesting  further improvement. We report on them
   and the  modifications for
 CRESST-II  in the present paper.

\section{Experimental Setup}\label{setup}
\begin{figure}\centering
\includegraphics[width=.5\textwidth]{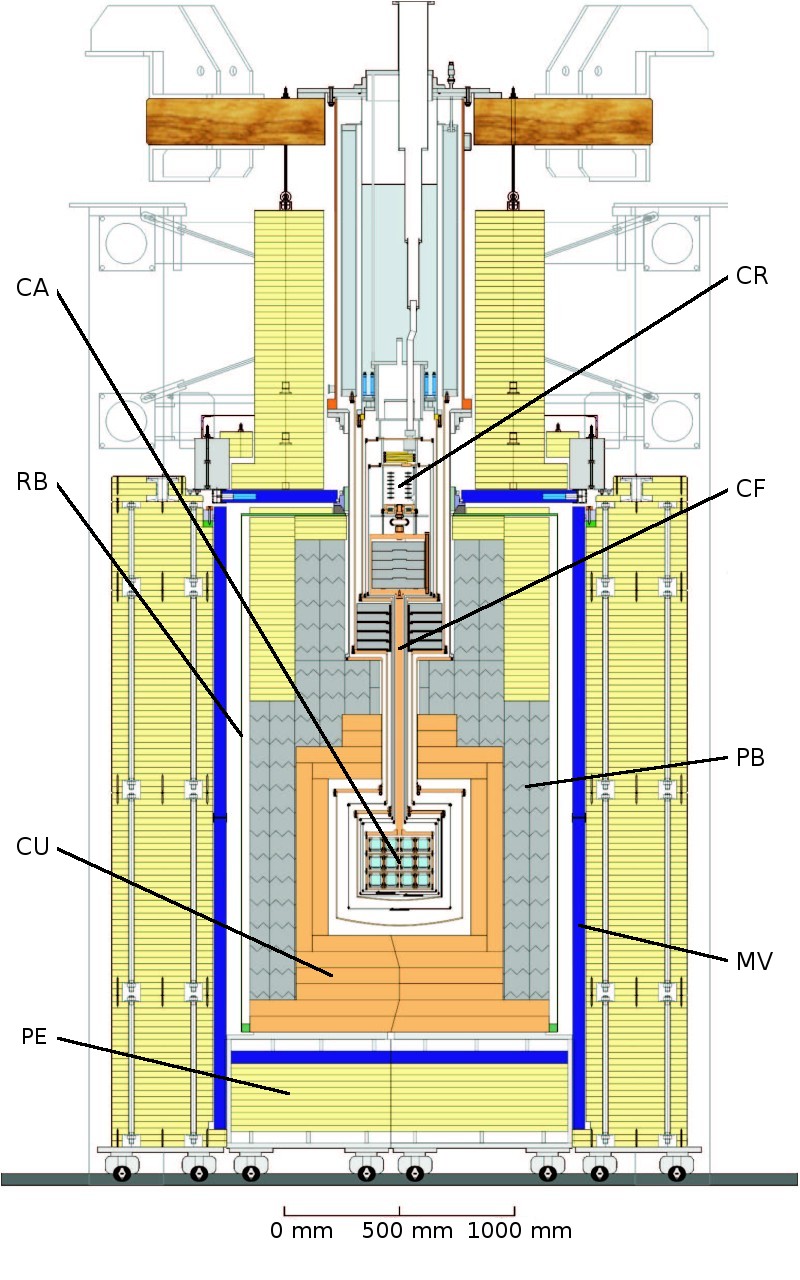}
\caption{$^3$He-$^4$He dilution refrigerator, low background
detector environment, and the shielding as upgraded for CRESST-II.
The detector carousel (CA) is connected to the mixing chamber of
the cryostat (CR)
by a long copper cold finger (CF) in order to reduce background 
originating from  the dilution refrigerator. The
gas-tight radon box (RB) encloses the low background copper
(CU) and low background lead  shielding (PB). It is covered by a
plastic scintillator muon-veto (MV) and a
$45\uu{cm}$ thick polyethylene
neutron moderator (PE). Additional granular
PE is placed
between the baffles in the upper part of the cryostat to close the
line of sight for neutrons coming  from 
above.}\label{fig:setup}
\end{figure}

The detector volume of the CRESST installation at Gran Sasso is 
surrounded by passive shielding consisting of   $14\uu{cm}$
thick low background copper and $20\uu{cm}$ of lead. This inner
shielding is entirely enclosed within a gas-tight radon box which
is continuously flushed with $\mathrm{N_2}$ gas. This part of the
setup is as used in previous phases of the experiment and is
described in Ref.~\cite{angloher2002}.
The changes  made in the  upgrade to CRESST-II
 include the installation of a neutron shield and a muon
veto. Figure~\ref{fig:setup} shows the setup after the upgrade.

In the detector volume itself, 
 a new detector support structure (``carousel") capable
of holding the 33 modules has been mounted and
a new 66-channel SQUID readout with associated wiring and data
acquisition electronics has been installed. The various changes
will be discussed in this section. 

\subsection{Detector Support Structure (``Carousel")}
\begin{figure}\centering
\includegraphics
[width=.5\textwidth]{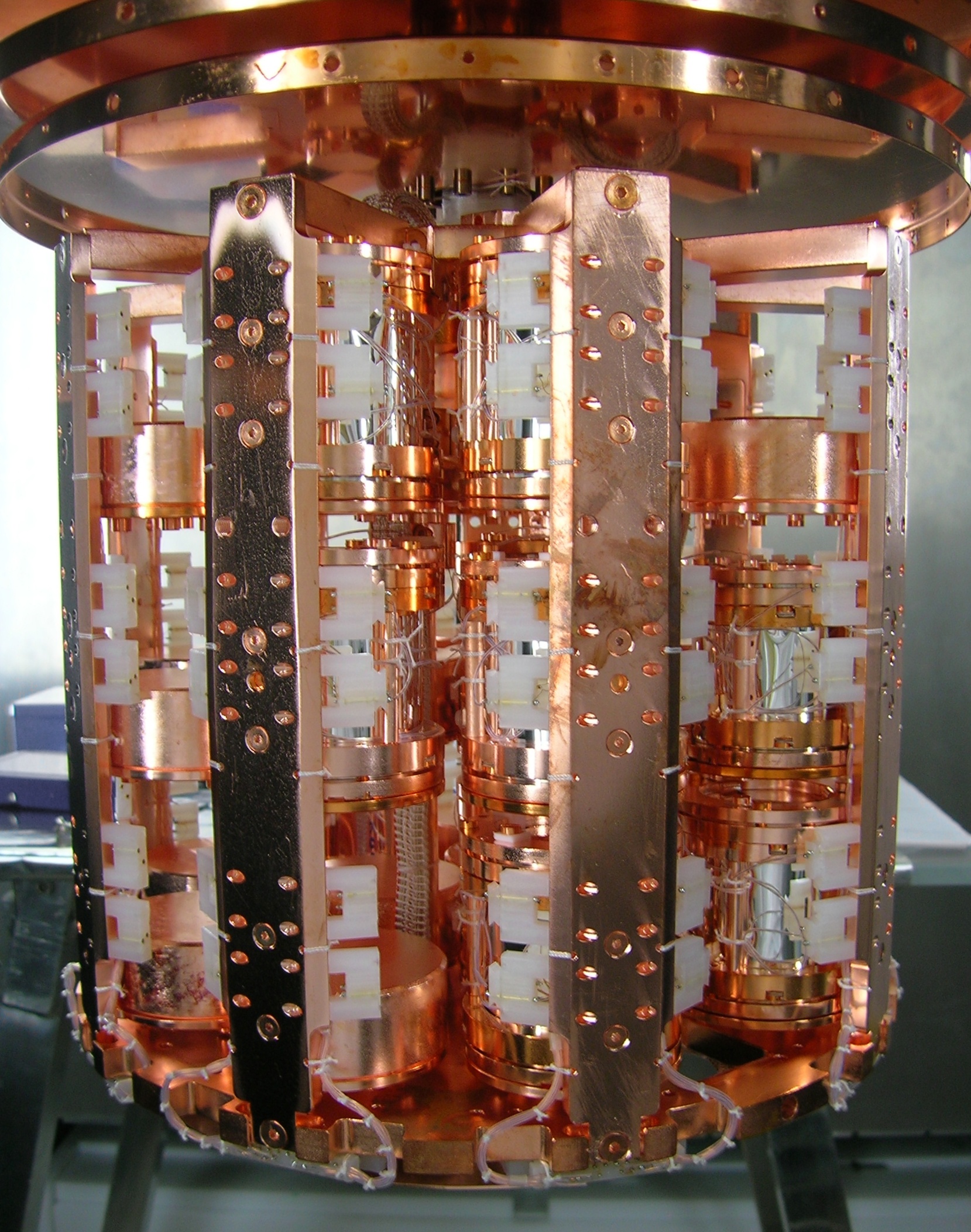}
\caption{ Detector carousel of ultrapure copper, 
electropolished to reduce  surface contamination. The  structure
can accommodate 33 detector modules (i.e. $10\uu{kg}$ of target
mass) which can be mounted or dismounted individually. The 
carousel is mounted at the lower end of the cold finger. The whole
structure is cooled to $\sim10\uu{mK}$.}
\label{fig:detectorsupporstructure}
\end{figure}

The new support structure for holding the  detector modules is
made from copper suitable for milliKelvin (mK) operation and 
electropolished to reduce surface contamination. The structure,
which  is depicted in Fig. \ref{fig:detectorsupporstructure}, can
accommodate a total of 33 detector modules, thus allowing an
increase of the target mass by an order of magnitude
(to $\sim10\uu{kg}$) relative to  CRESST-I. Individual handling and
dismounting 
of the modules  is possible.  The structure rests on custom-made
CuSn6 springs ~\cite{majorovits2008} to reduce interference
from vibrations. CuSn6 was chosen since it exhibits both 
intrinsic radiopurity and elasticity at mK temperatures. The
thermal coupling  of the carousel to the cold finger is 
intentionally set to be weak, with a relaxation time   of
about half an hour. This filters out high-frequency temperature
variations of the cryostat, giving  a high degree of 
temperature stability at the detectors.

\subsection{Sixty-six channel SQUID readout}

\begin{figure}\centering
\includegraphics[width=.3\textwidth]{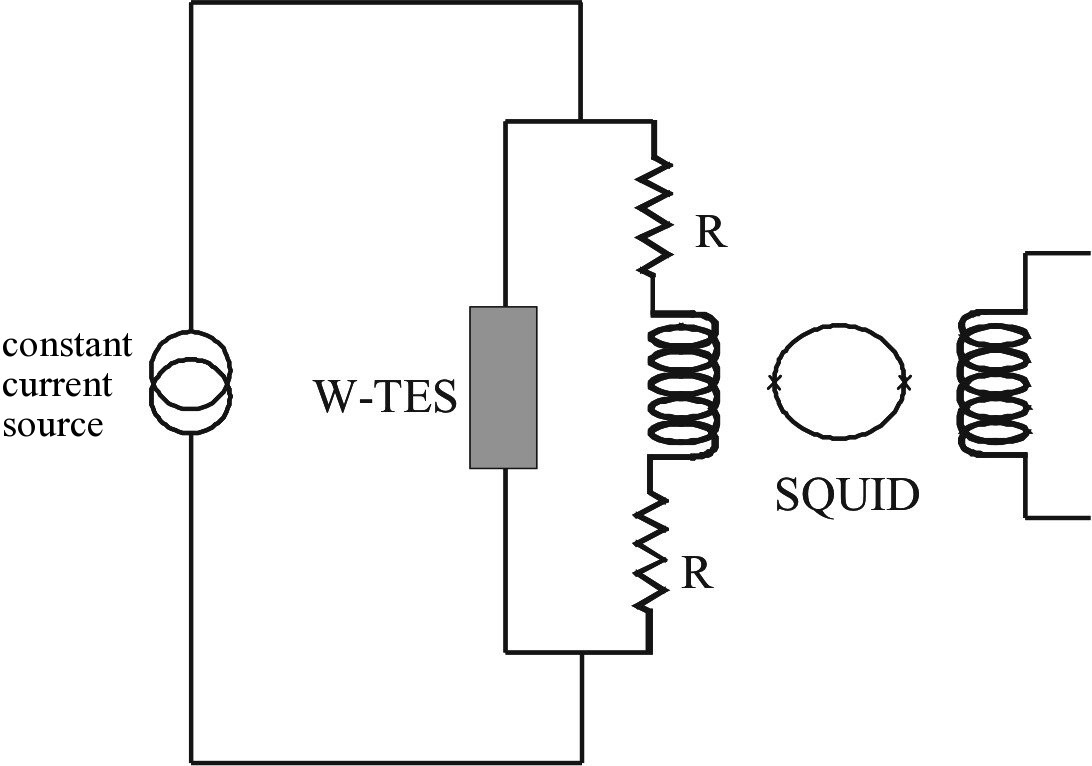}
\caption{ CRESST-II readout circuit. The
tungsten film thermometer is in parallel with  two reference
resistors and the  input coil of a  DC SQUID. The circuit is biased
by a constant current. 
}\label{fig:squid}
\end{figure}

With the full complement of 33 modules, each with
 two superconducting thermometers, there will be 66 readout
channels in CRESST-II. In the basic CRESST design these channels
are  read out with SQUID's. Figure \ref{fig:squid} shows one such
readout circuit, simplified for clarity. The actual implementation
is discussed in detail in Ref.\,\cite{henry2007}. A constant bias
current is shared between two branches, one containing
 the tungsten  thermometer, the other   the input coil for a  SQUID
in series with two
reference resistors. 
 The reference resistors 
assure proper
branching  of the current; two
are used here to make the circuit more balanced and to minimize the
level of crosstalk.  Any change in the resistance of the
thermometer changes the current through the SQUID input coil. The
SQUID system  then transforms this signal to a voltage pulse, which
is processed by conventional  follow-on  electronics. 
The SQUIDs are DC SQUIDs  using AC modulated flux-locked loops.
 Full details of the system are given in \,\cite{henry2007}.

To handle the large number of detectors, a 
multichannel system was installed, consisting of 66 SQUID
units. These are located  at the bottom of the main helium bath,
at 4 K.  Woven twisted pair
cables are used to connect the SQUIDs to the detectors in the mK
environment, while another set of cables leads to  connector
boxes at room temperature, which then link to the
control electronics and the data aquisition system. 

In addition to the readout-line pair for the SQUID, each
thermometer  has a heater-line pair for temperature control and
calibration pulses.
Thus there are two pairs of lines for each thermometer  and so  a
total of eight lines per module.
With the many  lines into the mK environment
 close attention must be paid in the design and fabrication of
flanges and feed-throughs \,\cite{henry2007}\cite{bela}.

\subsection{Neutron Shield}

A significant new element in the setup is a neutron shield, 
consisting of a
$45\uu{cm}$ thickness of  polyethylene (PE) plates. The shield is 
mounted
on rails to allow access to the inner parts of the experiment. The
 polyethylene   moderates ambient neutrons to thermal energies 
where they
are then captured on the hydrogen. The neutron capture range in 
$\uu{CH_2}$
materials is 4.3 cm for MeV energies~\cite{pfs}, so the shield
should eliminate essentially all incoming neutrons.

\subsection{Calibration Source Lift}

During a run, CRESST detectors are regularly calibrated with a
radioactive source supplying gamma quanta of known energy. In view
of the many and closely
spaced detectors, in CRESST-II it is necessary to be able to move
and position the  radioactive source  during calibration runs.

Therefore the apparatus has been equipped with a calibration source
lift.
The  lift consists of a plastic tube through which the source 
  is moved by  compressed
air, the  position being   determined by strings attached
from both ends of the tube. It
 spirals  around the cold box inside the shielding and  allows
calibration of 
individual detector modules without opening the shielding. It is
designed so that a rate of
$\sim0.5\,\uu{Hz}$  can be generated at any detector module  with
a 
0.7 MBq  $^{57}\uu{Co}$ source. The system was used successfully
several  times during the commissioning run.

\subsection{Muon Veto}
With the aim of reducing the effects of muons entering the
apparatus a muon veto has been  installed.
 It consists of 20 plastic scintillator panels  of area
about $1-1.2\uu{m^2}$  surrounding the radon box and
the Cu/Pb  shielding.  By using three different types of panels, a
total solid
angle coverage of 98.7~\% is achieved. A circular opening   of
$0.27\uu{m^{2}}$ on  
top  was unavoidable due to the entry for the
cryostat (see Fig. \ref{fig:setup}). In a dedicated setup
\cite{Nicolodi2005}, the characteristics of each panel type were
measured and  calibrated. 
The photomultiplier signals  are sent out of the Faraday cage
optically to 
 avoid interference with the cryogenic detectors. The 
veto triggered  at $\sim 5$ Hz during the run.

\section{Detectors}\label{detectors}

\subsection{ Module housing}\label{module}

An important feature of the CRESST detector modules is their
ability to reject  events which are not nuclear recoils via a
simultaneous measurement of a heat/phonon signal in the large
detector and a scintillation signal
\cite{meunier99} in the light detector. 
\begin{figure}\centering
\includegraphics[width=.5\textwidth]{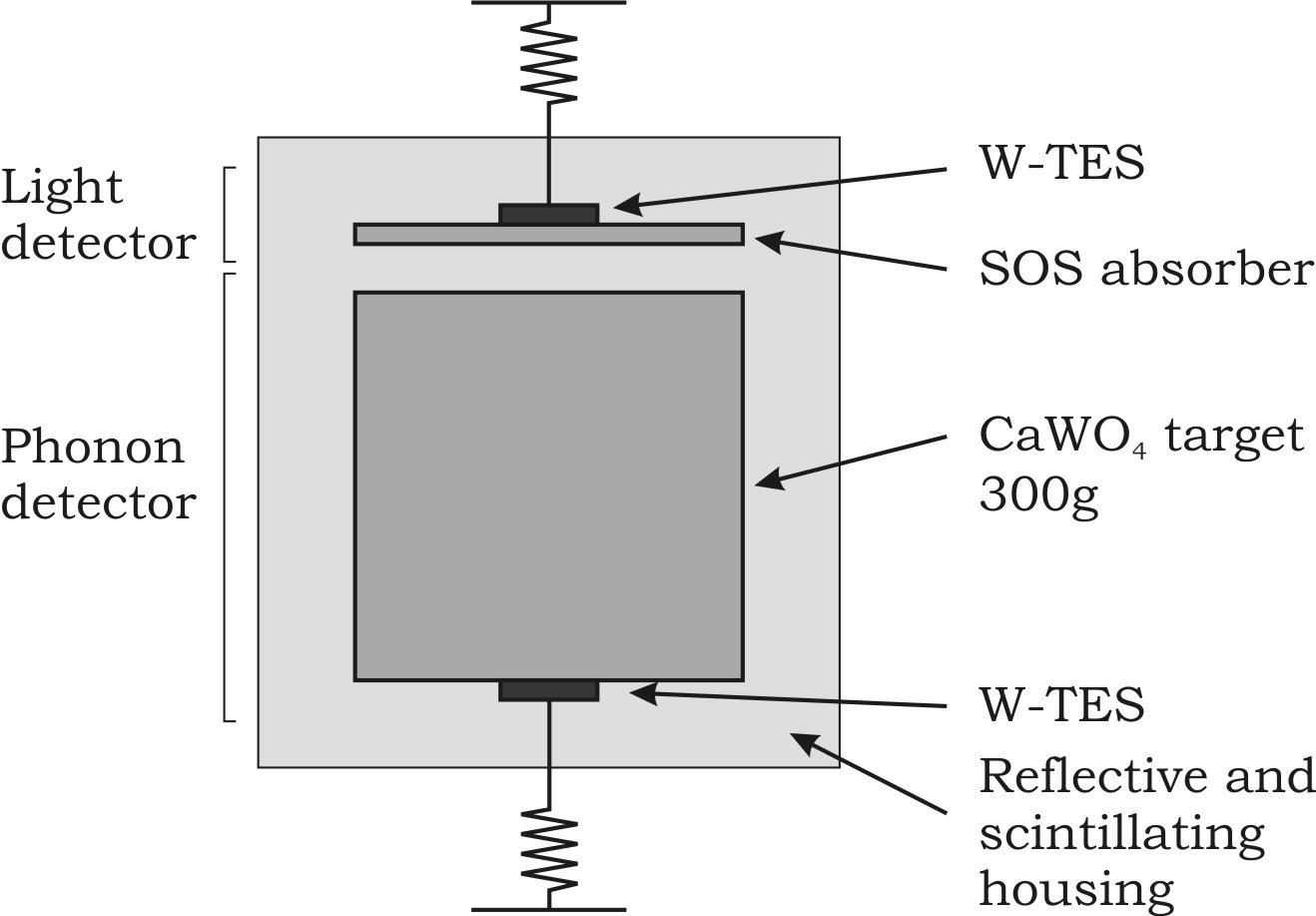}
\caption{Schematic arrangement of a CRESST detector module. The
module
consists of two independent detectors: one with the $\uu{CaWO_4}$
target crystal, providing a total energy measurement, and one with
a silicon-on-sapphire (SOS) wafer for measuring the scintillation
light   from the target crystal. Both detectors are enclosed in a
reflective, 
scintillating housing.}\label{fig:module}
\end{figure}

The crystal and the light detector are enclosed together in a
reflective and scintillating housing (Fig.\,\ref{fig:module}) of
 3M foil, as described in Ref.\,\cite{angloher2005}. This serves
two purposes. Firstly, it increases the amount of light collected
by the light detector. Secondly, its scintillation permits
rejection
of a potentially dangerous background from surface radioactivity 
on the crystal
 \cite{westphal2008}. An important such  possibility is
 polonium, a daughter
of the radon decay chain with an alpha decay. If the
alpha is
lost and the recoiling lead nucleus stays in the crystal,
this  can mimic a tungsten recoil, i.e. a WIMP
interaction.  However, it was found that the scintillation of the
alpha particle on the
reflecting foil gives a  light signal which discriminates
 events of this type and allows for their separation. 

\subsection{Thermometers (TES's)}\label{therm}
 The basic sensor in CRESST is a superconducting thin film
thermometer, with the temperature 
stabilized at a suitable operating point in the middle of the
superconducting-to-normal transition of the  film . A small
variation of the film resistance in such an arrangment gives a very
sensitive temperature measurement; this system
 is often called a Transition Edge Sensor (TES) .

In the work described here
 tungsten thin films are used.
 Tungsten in its crystalline $\alpha$ phase
becomes superconducting
at $T_c\approx15\uu{mK}$.
The films are equipped with heaters with the dual function of
  stabilizing the detectors' operating points
and the injection of heat pulses for calibration.
The thermometer layouts for the $\uu{CaWO_4}$ target and the light
detector   are
similar.   However, the light detector has additional  Al/W phonon
collectors for better energy collection. This is as
described in ref.~\cite{angloher2005}, but here  the
light detectors use
a small portion of the thermal-link gold structure as the heater.
This light detector layout  is shown in  Fig.
\ref{fig:lightdetectordesign}.

 For 
temperature regulation
heater  pulses of a fixed amplitude,
so-called control pulses are injected through the heater lines
every three seconds. Online
pulse height evaluation of the resulting signal is used for 
regulation of the heater current which keeps the thermometer's
temperature  at
the operating point. This is somewhat different from older work
where temperature stabilization was on the baseline, and 
 is found to provide  excellent long term stability of operation.
In addition
to these control pulses for temperature stabilization,  calibration
pulses delivering a known energy  are  injected every
30 seconds. These are used in determining the energy to be
identified with a pulse, as will be explained in section \ref{ed}.
The  calibration and stabilization procedures are the same for both
target and light detector thermometers.
\subsection{Target Detector}\label{phonon}

The  CRESST-II detectors use large $\mathrm{CaWO_4}$ single
crystals as the  target or ``absorber'' mass.
$\uu{CaWO_4}$ has been selected for its high light output and the
presence of the heavy  nucleus  tungsten, which  gives a large
factor $\sim A^2$ in the presumed coherent (spin-independent)
scattering of the
WIMP. 
The  crystals   are cylindrical, of
$4\uu{cm}$ height and  $4\uu{cm}$ diameter, and  about
$300\uu{g}$ in weight. 

  Ten such
crystals were installed for the commissioning run;
 two are used here for the dark matter analysis.

\subsection{Light Detector}\label{light}

\begin{figure}\centering
\includegraphics[width=.5\textwidth]{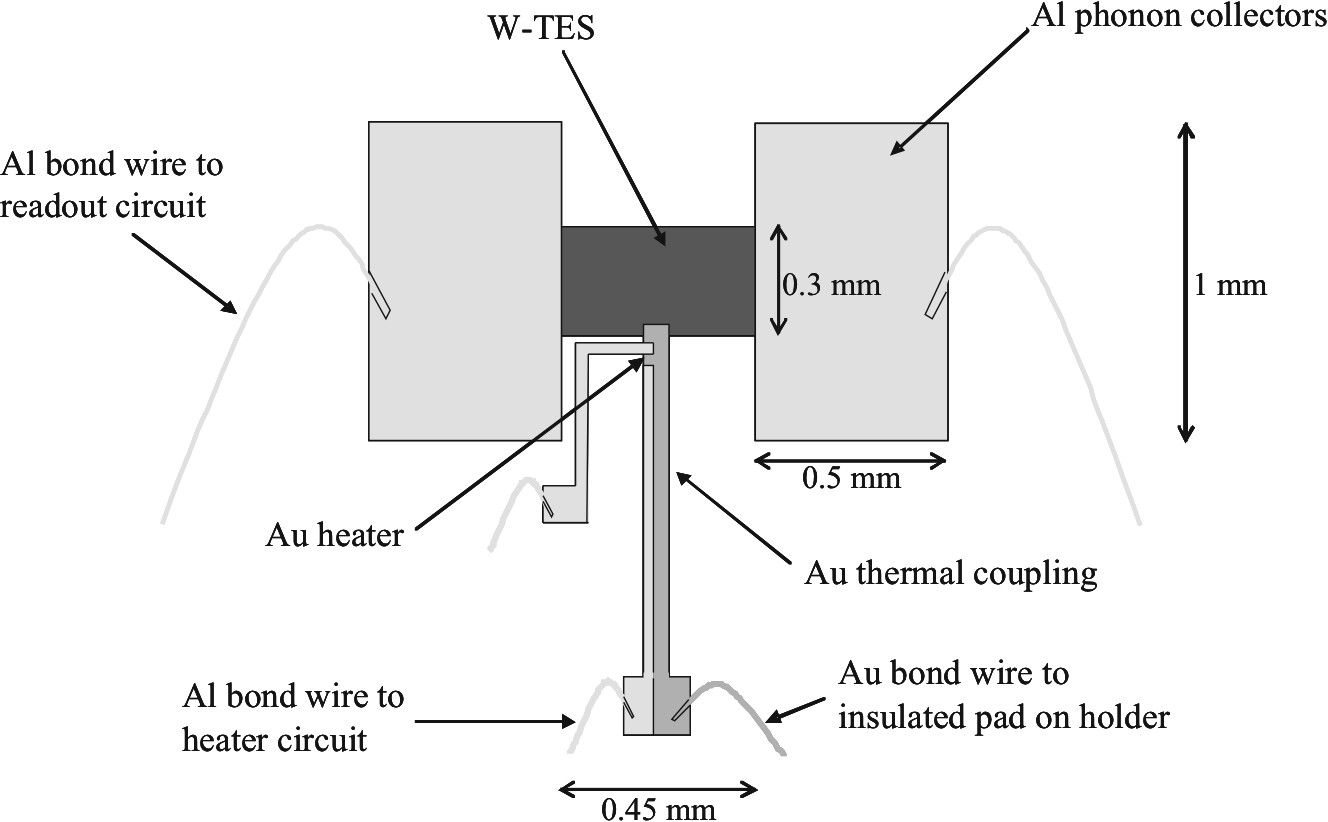}
\caption{Layout  and connection scheme of a light detector used in
the commissioning run. Aluminum/tungsten phonon collectors surround
the tungsten
thermometer or TES. A small portion of the gold thermal link is
used as the heater. Connections shown are the aluminum bond wires
for  the SQUID and heater circuits, and a gold bond wire   for 
thermal coupling, leading to an electrically insulated pad on the
detector holder.}\label{fig:lightdetectordesign}
\end{figure}

The scintillation light  is measured by a separate cryogenic
detector. This is made from a sapphire wafer of $40\uu{mm}$ 
diameter and
$0.4\uu{mm}$ thickness, with an epitaxially grown silicon
 layer on one side for  photon absorption. The weight is 2.3 g.
This
silicon-on-sapphire (SOS)
material  exploits the excellent phonon transport
properties of sapphire.  The resulting  detector with the sensor as
in Fig. \ref{fig:lightdetectordesign} is sensitive    down to about
20 eV, or about seven optical
photons.

 Further progress in using the light yield to identify which
nucleus is recoiling,
as discussed below,
depends largely on the improvement of the resolution of the light
detector.

\section{Data Taking and Analysis}\label{data}

Only very basic quality cuts are performed on the data. Pulses are
rejected if they occur in one of the rare periods when the
temperature of the cryostat is not  stable, as
inferred from the pulse height of the control pulses.
Pileup is rejected by simple cuts on the baseline of the
pulse. Due to the low trigger rate this  does not
introduce a significant dead time into the analysis.

\subsection{Signal Processing and Pulse Fitting}
For recording  thermometer pulses, the output voltage
of the SQUID electronics is split into two branches. In one branch
the pulse is shaped and AC-coupled to a trigger unit, while in the
other branch the signal is passed through an 8-pole anti-aliasing
low-pass
filter and then DC-coupled to a 16-bit transient digitizer. The
time bin
 of the transient digitizer was chosen to be $40\uu{\mu s}$,
providing about 30 samples for the rising part of the pulse in the
phonon channel. The record length of 4096 time bins includes a
pre-trigger region of 1024 samples to record the baseline before
the event, while the remaining 3072 post-trigger samples contain
the pulse itself. The phonon and light channels of each 
module are read out together, whenever one or both trigger.

The pulses are fitted offline using a template fit, with a template
constructed from pulses collected during  a gamma calibration run.
There is an individual template for each thermometer. Both pulses
from a module are fitted simultaneously. Free parameters in
the fit are the two baselines, the two pulse height amplitudes, and
a common time shift relative to the trigger.

\subsection{Energy  Determination}\label{ed}
 Finally, the assignment of an energy to a pulse is made on
the basis of the resulting  amplitude or pulse height parameter.
Since the  pulses
from the  $^{57}\uu{Co}$ calilbration  with $\sim 122\uu{keV}$ and
$\sim 136\uu{keV}$  photons
establish a relation between pulse height and energy for each
detector, a first simple determination of the pulse height-energy
relation for a detector would be provided by a linear extrapolation
from $ 122\uu{ keV}$ to lower energies, and in early work this gave
good
results. Direct calibration with lower energy gammas is not
possible since the calibration source is outside the coldbox and
the photon must penetrate about $12\uu{mm}$ of
copper. 

 A finer and more detailed
determination of the pulse height-energy relation for a given
detector  is possible, however, by using the heater calibration
pulses mentioned in section\,\ref{therm}. For each thermometer,
heater
pulses   delivering a known energy via ohmic heating   are 
injected every
30 seconds. These are applied over the whole energy range of
interest, so the thermometer response can be mapped down to low
energies and the assumption of a simple linear extrapolation
can be avoided.
 Both target and light detectors are calibrated by
this procedure.

\begin{figure}\centering
\includegraphics[width=.5\textwidth]{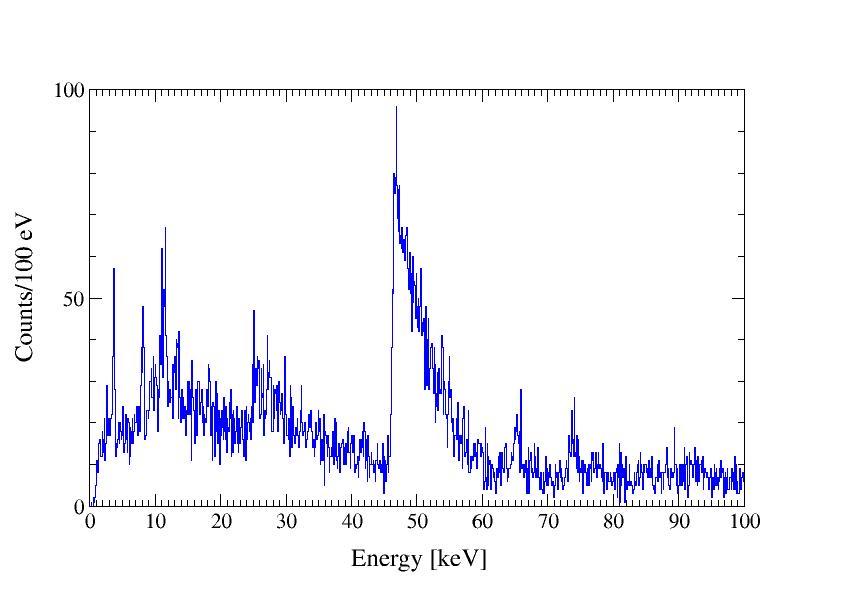}
\caption{ Energy spectrum of the 300g detector ``Verena" in the
commissioning run (48 days). The 
prominent feature at $46.5\uu{keV}$ is due to a gamma from 
$^{210}Pb$ decay in the crystal with its associated $17\uu{keV}$
$\beta$ spectrum. The
Cu fluorescence line at $8.05\uu{keV}$ is also clearly visible, and
has resolution $\sim300\uu{eV}$. The close agreement of these
energies with known values
confirms the excellent accuracy of the  
heater pulse calibration method.}
\label{fig:energyspectrumverena}
\end{figure}

The accuracy and resolution of  this
method is illustrated in
 Fig.
\ref{fig:energyspectrumverena}, which shows the energy spectrum of
the target detector ``Verena" in the commissioning run.  The energy
resolution may be gauged from the  prominent feature starting
around $46\uu{keV}$ and the line near $8\uu{keV}$, which can be
identified as a copper fluorescence line. The large feature   is
due to the gamma  transition of a  $^{210}\uu{Pb}$
impurity in the crystal, with a broadened right shoulder from an
accompanying $17\uu{keV}$ $\beta$ spectrum. The steep left edge
appears precisely
at $46.5\uu{keV}$ as expected. The peak for the copper
fluorescence line  is at
$8.05\uu{keV}$, also as expected, and the energy resolution is
$\sim300\uu{eV}$. These two features confirm the excellent accuracy
of
the calibration method with heater pulses. Further details of the
energy spectrum can also be understood and will be discussed in a
forthcoming publication.

\subsection{Energy-Light  Plots and Quenching Factor}\label{elp}

Events from a given module are plotted as points in the
 energy-light yield plane, as  in the following Figs.
\ref{fig:ncal} and
\ref{fig:background}.  An energy is assigned to an
event  as described in the previous section. The light yield
is given by the ratio of the energy  or pulse height in the light
detector to that in the target detector, normalized such that 
it is  1 for
 a 122 keV calibration photon absorbed in the large detector. While
the light  yield  for a 122 keV photon  is therefore one by
definition, the fact
that  the bands are roughly flat on the plots,
and the average light yield
is always near one, is not a matter of definition and reflects the
accuracy of the thermometer calibrations and  the  approximate
linearity\footnote{Upon closer examination
it appears that the centroid  of the electron-photon band falls  
slightly at lower energies. 
Such a decrease for $\gamma$ rays in ${\mathrm {CaWO_4}}$
 has been seen  in Ref.\,\cite{flat}.} of the light production
process~\cite{bavykina2007}.

It will be seen that when a source inducing nuclear recoils is
present, as with neutrons in 
Fig.\,\ref{fig:ncal}, there are two distinct bands, one of lower
light yield from nuclear recoils and one of higher light
yield from electron-photon events. The factor of reduction for
 the light yield  of a  nuclear recoil relative to that of the
electron-photon event of the same energy is called (in our usage)
the quenching
factor. With the
electron-photon band centered on 1, the center of the nuclear
recoil band on Fig.\,\ref{fig:ncal} is  simply at the inverse of
the
quenching factor.

Since the quenching factor  varies widely, up to 
$\sim 40$ for heavy elements, the level of light output can be used
not only to distinguish nuclear recoils from electron-photon
events, but also to distinguish tungsten recoils from those
of the lighter nuclei. It is hoped that it may be ultimately
 possible to
determine which particular nucleus, in a compound material like
$\mathrm{CaWO_4}$, is recoiling. This would be very useful in
verifying a
positive dark matter signal, since a simple behavior for the signal
with
respect to changing the target nucleus is expected.

Since an understanding of the quenching factor  is thus of  great
interest, a number of investigations have been carried out within
the CRESST collaboration
 \cite{westphal2008},\cite{mikhailik2006},\cite{ninkovic2006},
\cite{jagemann2006},\cite{coppi2006}. The results are summarized
and discussed in~\cite{bavykina2007}.

\section{Neutron  Test} \label{nct}
To directly test the response of the system to nuclear recoils
a neutron test was carried out during the commissioning
run.  The test can  be used to 
 check the principle of   identifying  the recoiling nucleus  via
the light yield and   to also determine the dispersion in the light
output, which is important in applying  the method.

Above $\sim 10\uu{keV}$ recoil energy, calculations show that the 
spectrum for neutron scattering on $\uu{CaWO_4}$ is totally 
dominated by oxygen recoils \cite{privwu}.
Therefore one anticipates that the events of the nuclear recoil
band in this test should  lie roughly  on or below the line
corresponding to
the oxygen quenching factor  $\sim 9$ known from the previous
quenching factor studies. 

\begin{figure}\centering
\includegraphics[width=.45\textwidth]{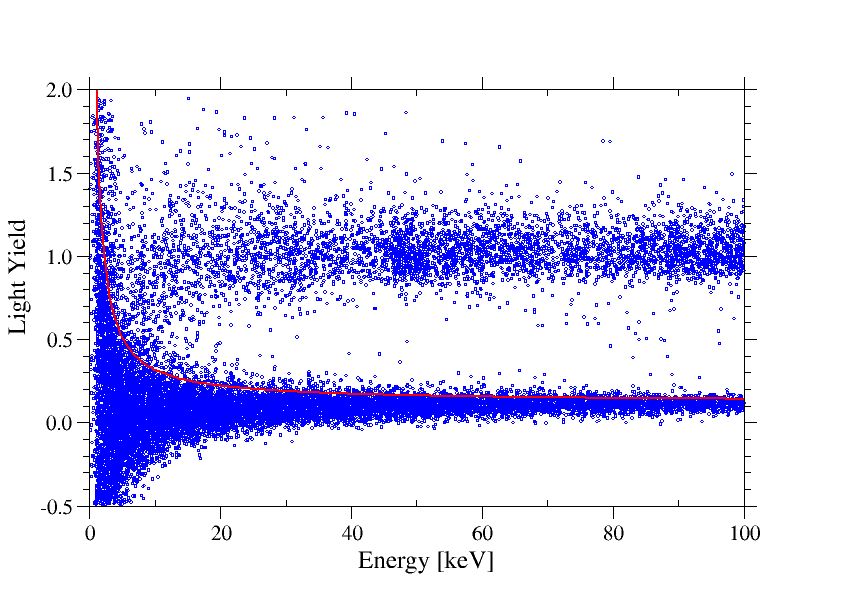}
\caption{Low energy event distribution  in the (Verena/SOS21)
module
in the presence of a neutron source during the commissioning run.
Below the red
curve 90\,\% of the oxygen recoils are expected, as calculated
from the known quenching factor for oxygen and the  energy
resolution of the light detector. The good agreement with the
prediction supports the use of this method to identify
tungsten recoils in the dark matter analysis. The events above
$47\uu{keV}$ between the electron and
nuclear recoil bands can be understood as  inelastic neutron
scatterings where the tungsten nucleus is excited to a
gamma-emitting
level.}\label{fig:ncal}
\end{figure}

A neutron source was placed inside the polyethylene shield and 
Fig.~\ref{fig:ncal} shows the results for one of
the two modules (Verena/SOS21). Using the  quenching factor of 9 
and the measured energy-dependent resolution  of the light
detector, 90~\% of
the   neutron-induced
recoils should be below  the red line.

  The
neutron test data  appears to  agree  well with this prediction,
and
so supports our   arguments in the dark matter analysis that an
acceptance region for nuclear or  tungsten recoils based on the
same principle can be defined.

\section{Dark Matter Limits} \label{results}

\begin{figure}\centering
\includegraphics[width=.5\textwidth]{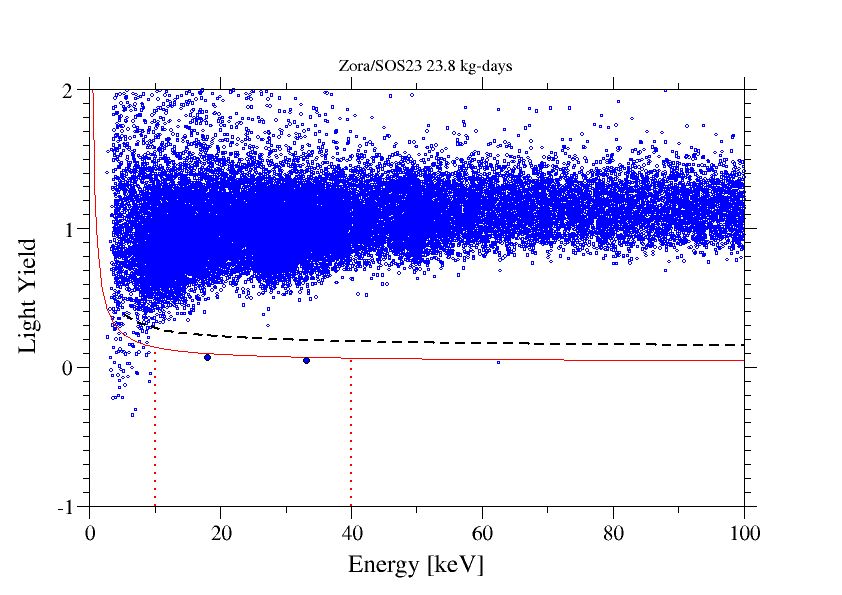}
\vspace{0.5cm}
\includegraphics[width=.5\textwidth]{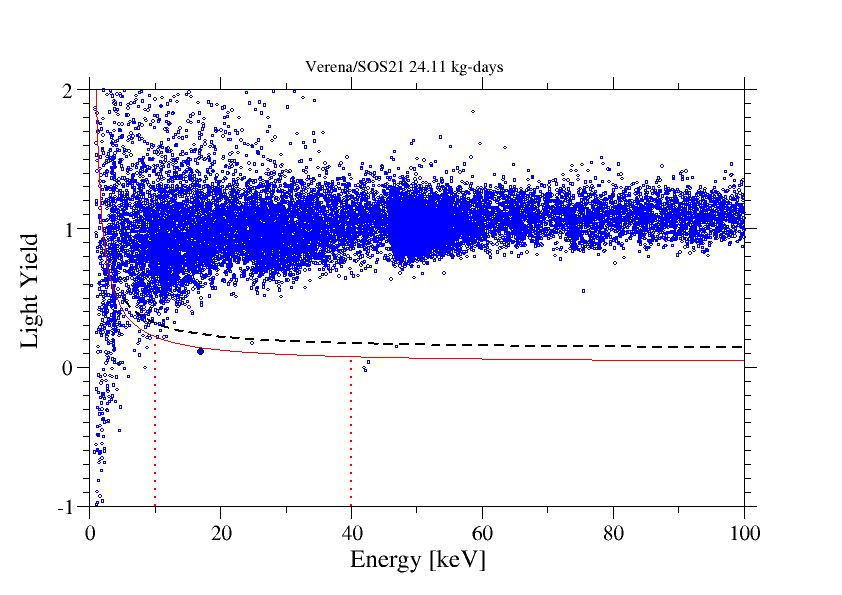}
\caption{Low-energy event distribution measured with two
$300\uu{g}$ $\mathrm{CaWO_4}$ detector modules during the
commissioning run. The vertical axis represents the light yield
(see text), and the horizontal axis  the total energy, as
measured by the phonon channel. Below the dashed black curve 90\,\%
of all nuclear recoils, and below the solid red curve 90\,\% of the
tungsten recoils are expected. The heavy black dots show the events
in the ``tungsten recoils'' acceptance region. The intense regions
of the electron recoil bands correspond to the lines of  Fig.
\ref{fig:energyspectrumverena}.}
\label{fig:background}
\end{figure}

For the  dark matter analysis we use
 data  taken 
between March 27th and July 23rd  2007 with the   two detector
 modules (Zora/SOS23) and (Verena/SOS21). The data are shown
in Fig.~\ref{fig:background}.
The cumulative exposure was $47.9\uu{kg-days}$. For the 
    tungsten  this corresponds to an exposure of
$30.6\uu{kg-days}$.

We perform an analysis on the assumption of coherent or
 spin-independent scattering for the WIMP. This process should
strongly
favor tungsten recoils due to the $\sim A^2$ factor in the WIMP-
nucleus cross section.
 We define an acceptance region on the plots based on a)  the
quenching factor for the light yield  and  b) the 
maximum energy expected for tungsten recoil. A  similar
region for ``all nuclear recoils'' can be defined using
the quenching factors for Ca and O.
The quenching factor boundaries are shown on the plots
of Fig.~\ref{fig:background} by the curves. Below the upper curve
90\,\% of all
nuclear recoils are expected, and below the lower curve 90\,\% of
the tungsten recoils are expected.
The   energy boundaries are shown by the vertical lines. The upper
limit at  $40\uu{keV}$ is set by  form-factor\,\cite{helm1956}
effects, which effectively limit the energy transfer to the
tungsten nucleus. The
lower limit is  set at $10\uu{keV}$, where ``leakage'' from the
electron/photon events becomes evident and so recoil discrimination
becomes inefficient.

\begin{figure}\centering
\includegraphics[width=.5\textwidth]{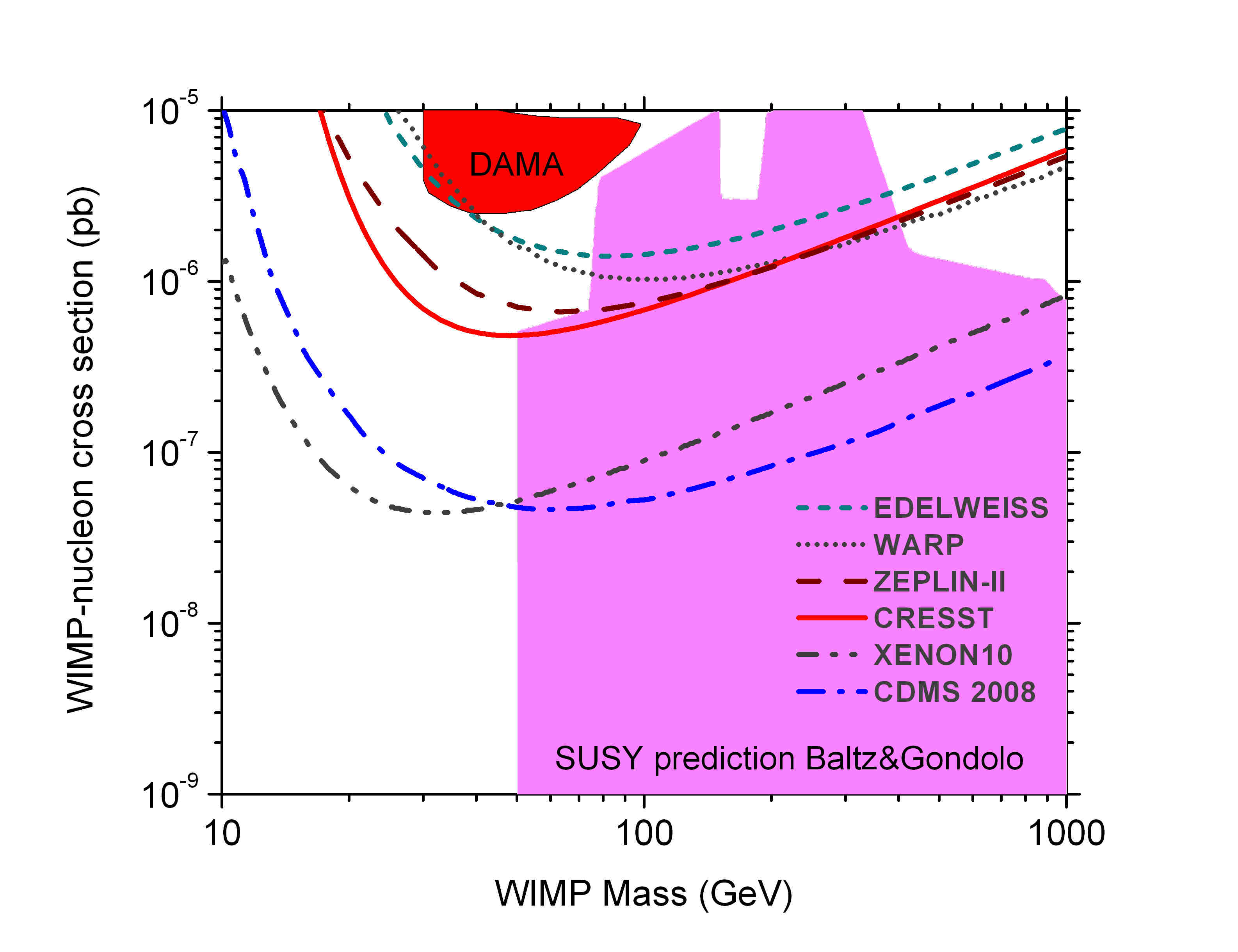}
\caption{Coherent or spin-independent scattering cross section
exclusion limit
derived from the data of Fig.~\ref{fig:background} using the
maximum energy gap method. For comparison the limits from other
experiments \cite{cdms08},\cite{xenon10},\cite{warp},
\cite{edelw},\cite{zep}
 and the range predicted by some
supersymmetry models~\cite{balz2000} are also
shown.}\label{fig:limit}
\end{figure}

Three candidate events (heavy black dots) are observed in
Fig.~\ref{fig:background} for the ``tungsten recoils'' acceptance
region.
The individual events are at 
16.89\,keV (Verena/SOS21)  and  at  18.03\,keV and 33.09\,keV
(Zora/SOS23).
 The  corresponding  rate is 0.063  per kg-day.
From this rate we can
 derive a limit on the coherent spin-independent WIMP-nucleon
scattering cross section. Using standard assumptions on the dark
matter
halo\,\cite{donato1998,lewin1996} (WIMP mass density
of $0.3\uu{GeVcm^{-3}}$),   an upper limit for the coherent or
spin-independent WIMP-nucleon scattering cross-section may be
obtained using the maximum energy gap method~\cite{yellin2002}.
This limit is
plotted as the red
curve in Fig.~\ref{fig:limit}. The minimum of the curve, for a
WIMP mass of $\sim$50 GeV, is at $4.8\times10^{-7}\uu{pb}$.

\section{Backgrounds}
To judge the improvement with respect to our previous setup of
Ref.\,\cite{angloher2005} we take the ``all nuclear recoils''
region
and  reset the low energy boundary of the acceptance region to
12\,keV as it was there. In this acceptance region the earlier work
  had 16 events, for a  rate of  0.87 per kg-day. As noted in
Ref.\,\cite{angloher2005}, this rate was compatible with the
estimates of \cite{wulandari2003} for neutron background.
 The present commissioning run  has 4 events in the same acceptance
region,
 for  a rate of 0.083
per kg-day, so we may say there is a factor $\sim 10$
improvement.  We attribute this 
to the presence of the neutron shield. 

 On the other hand,   according to the arguments in 
section\,\ref{nct}
the ``tungsten recoils'' should be  insensitive to  neutron
background. Indeed, the three events here from 48 kg-days  are
quite compatible with the zero events
from 10 kg-days  in
Ref.\,\cite{angloher2005}; the present rate   of 0.063
per kg-day
 predicts only 0.6 events for 10 kg-days. The fact that the neutron
shield appears to have a great effect on ``all nuclear recoils''
but little effect on ``tungsten recoils''
 supports  our argument that the light yield successfully
distinguishes tungsten and other nuclear recoils.

The question arises, however, as to the nature of the few observed
tungsten
or nuclear recoil candidates.
 One possibility is evidently remaining  neutrons. However  with 
many absorption lengths for  the shield very few should penetrate 
and indeed
simulations  \cite{Wu04b} would give a rate of only
$\sim 10^{-5}$ per kg-day, much less than the few events. 
A possible point here is that
 during the run a weak spot in the neutron
shielding above the muon veto was identified and  patched only
after data taking was completed.

 Another  possible source for neutrons  are some non-operational
modules
which were present during the run. These can act as
non-vetoable neutron sources for the operational modules. However
an estimate for the
background events from an inactive  detector module is
well below $1.4\times 10^{-5}$ per kg-day.  

Apart from neutrons, another possibility arises from incomplete
coverage of the inner surfaces of the detector module with
scintillator, which could lead to unvetoed
nuclear recoils from surface $\alpha$-decays as discussed in
section \ref{module}. In further work it is intended to paint some
possibly uncovered areas with scintillating material.

Concerning muons,
 estimates\,\cite{Wu04b} of  muon-induced neutrons in the setup
result in only $2.8\times10^{-3}$ per kg-day. No muon signals
were found in coincidence with  nuclear recoil events.

 There thus appears to be no conclusive explanation for the few
candidate events from conventional  radioactive  or particle 
sources. We hope to clarify some of these points  in further work.

\section{Conclusions}\label{summary}

CRESST-II successfully completed its commissioning run in 2007.
New elements of the apparatus  were successfully installed and
operated. A neutron test demonstrated the ability to detect nuclear
recoils with a light yield consistent with that found from
quenching factor studies, supporting the principle of identifying
the recoil nucleus via the light yield.

Data were taken with two detector modules  for a total of 48
kg-days.
Three candidate events of uncertain origin are present in the
acceptance region for  tungsten recoils, yielding  a rate of 0.063
per kg-day.
A factor $\sim 10$ improved performance is found with respect to
previous work for the ``all nuclear recoils'' acceptance region.  

A limit on  coherent WIMP-nucleon scattering, 
is obtained, which at its lowest  value, for $M_{\uu{WIMP}}\approx
50 $GeV, is $4.8\times10^{-7}\uu{pb}$.

\section{Acknowledgements}\label{acknowledgements}

This article is dedicated to the memory of our friend and colleague
Wolfgang Westphal, who  significantly contributed to the success of
CRESST.

This work was partially supported by funds of the DFG (SFB 375,
Transregio 27:``Neutrinos and Beyond"), the Munich Cluster of
Excellence (``Origin and
Structure of the Universe"), the EU networks for Cryogenic
Detectors
(ERB-FMRXCT980167) and for Applied Cryogenic Detectors (HPRN-
CT2002-00322),
and the Maier-Leibnitz-Laboratorium (Garching).  We  also  thank
Hans Ulrich Friebel for valuable technical support.

 Support was provided by the Science and Technology Facilities
Council.

\end{document}